\documentclass[aps,twocolumn,showpacs,superscriptaddress,nofootinbib]{revtex4}

\usepackage[dvips]{graphicx}
\usepackage{dcolumn}
\usepackage{bm}

\usepackage{young}
\newdimen\hoogte    \hoogte=6pt
\newdimen\breedte   \breedte=8pt

\begin{document}

\title{P-wave Pentaquark and its Decay in the Quark Model\\ with Instanton Induced Interaction}

\author{Tetsuya Shinozaki}%
\email[e-mail: ]{shinozk@th.phys.titech.ac.jp}
\affiliation{%
Department of Physics, H-27, Tokyo Institute of Technology,\\ Meguro, Tokyo 152-8551, Japan
}%
\author{Makoto Oka}%
\affiliation{%
Department of Physics, H-27, Tokyo Institute of Technology,\\ Meguro, Tokyo 152-8551, Japan
}%
\author{Sachiko Takeuchi}%
\affiliation{%
Japan College of Social Work, Kiyose 204-8555, Japan
}%

\begin{abstract}
$P$-wave pentaquarks with strangeness $+1$, $I=0$ and $J^P=1/2^+$ are studied in the non-relativistic quark model with instanton induced interaction (III). 
We present their mass splittings and orbital-spin-isospin-color structures.
It is found that decompositions of the wave functions are sensitive to III, while the mass splittings are insensitive.
The decay of the lowest energy pentaquark, $\Theta^+$, is found to be suppressed when the contribution of III is increased. Its wave function is dominated by Jaffe-Wilczek-type configuration at large III.
\end{abstract}

\pacs{14.20.-c, 12.39.Mk, 12.39.Jh, 12.38.Lg}

\maketitle

\section{\label{sec:Introduction}Introduction}
$\Theta^+(1540)$ was discovered in 2003 at SPring-8~\cite{Nakano:2003qx}.
Follow-up experiments and reanalyses of old data claimed to confirm the existence of $\Theta^+$.
On the other hand, there have appeared some experimental results which show no signs of $\Theta^+$.
Until now, the existence of $\Theta^+$ is controversial~\cite{Danilov:2005kt}.
Peaks corresponding to $\Phi^{--}$ and $\Theta^+_c$ were also reported but still unconvincing.
The experimental results of $\Theta^+$ show that 
the mass is about 1540MeV, the upper limit of the width is a few MeV, the isospin $I=0$, the baryon number $B=+1$ and the strangeness $S=+1$. 
Since the minimal quark component of $\Theta^+$ is $uudd\bar{s}$, $\Theta^+$ is called ``pentaquark''.
The spin and the parity of $\Theta^+$ have not been determined yet.
In this work, we consider the case in which the spin and the parity of $\Theta^+$ are $J^P=1/2^+$~\cite{Diakonov:1997mm,Jaffe:2003sg}.

The mass of $\Theta^+$ has been investigated based on the quark model~\cite{Jennings:2003wz,Bijker:2003pm,Carlson:2003pn,Karliner:2003dt,Kochelev:2004nd,Kanada-Enyo:2004bn,Takeuchi:2004fv,Stancu:2004du,Hiyama:2005cf,Dmitrasinovic:2005gq}, the QCD sum rule~\cite{Sugiyama:2003zk,Nishikawa:2004tk,Kondo:2004cr,Wang:2005pq} and the lattice QCD~\cite{Sasaki:2003gi,Csikor:2003ng,Mathur:2004jr,Ishii:2004qe,Ishii:2005vc,Takahashi:2005uk,Lasscock:2005kx}.
Most of them predict the mass is much larger than the observed one $\simeq1540$MeV.
We, however, employ non-relativistic quark model in this work and do not attempt to reproduce absolute masses of pentaquarks.
In the non-relativistic quark model, absolute masses of ordinary meson and baryons are often adjusted to the ground state.
The mass splittings come mainly from the hyperfine interactions, $e.g.$ one-gluon exchange interaction and instanton induced interaction (III).

The observed width of $\Theta^+$ is unexpectedly narrow,  considering that the decay of $\Theta^+$ requires no pair creation.
Attempts have been made to explain the width based on the quark model~\cite{Jaffe:2004at,Hosaka:2004bn,Melikhov:2004qh} and the QCD sum rule~\cite{Ioffe:2004qm,Eidemuller:2005jm,Wang:2005ms,Melikhov:2004ws}.
However, most of them suggest that the centrifugal barrier and the symmetry properties of the orbital-spin-flavor-color wave function can not make the width as narrow as a few MeV.
Furthermore, such choices of the wave function, which are expected to give a small width, seem quite artificial.
In this work, we point out that the instanton plays an important role, and III explains why such choices are favorable.
As a result, III is shown to make the width of $\Theta^+$ narrower.

The instanton is a classical solution of the Yang-Mills equation in the Euclidean space~\cite{Belavin:1975fg}, which is one of the most important non-perturbative effects in QCD. 
Importance of instantons in low-energy hadron phenomena can be seen from the $U_A(1)$ symmetry breaking.
It is well known that the large mass of $\eta'$ indicates that the $U_A(1)$ is broken due to anomaly but not by spontaneous symmetry breaking~\cite{Weinberg:1975ui}.
It is suggested that the anomaly comes from the instanton in the QCD vacuum~\cite{'tHooft:1976fv,'tHooft:1976up}.
Instanton induced interaction (III) is an effective inter-quark interaction through the zero modes of the light quarks around an instanton~\cite{Shifman:1979uw}.
III makes the mass of $\eta'$ heavy, thus reproduces the low-lying meson spectrum.
Effects of III for the mass of $\Theta^+$ have been investigated~\cite{Shuryak:2003zi,Kochelev:2004nd,Wang:2005pq}, but they study only limited cases. We treat all the $P$-wave states and study the decay of $\Theta^+$ also.

In section~\ref{sec:Formalism}, we show our quark model Hamiltonian and enumerate all the $P$-wave states and discuss the connections between the hyperfine interactions and the decay widths.
In section~\ref{sec:Results}, we show the results of the pentaquark masses and widths.
In section~\ref{sec:Conclusion}, we give a conclusion and an outlook.

\section{\label{sec:Formalism}Formalism}

\subsection{Hamiltonian}
Hamiltonian of the non-relativistic quark model~\cite{Isgur:1978xj} is given by
\begin{equation}
H = M_0 + \sum_i \frac{\vec{p_i}^2}{2m_i} + V_{conf} + H_{HF},\label{eq:Hamiltonian}
\end{equation}
where $M_0$ is a constant term. $m_{i}$ is the constituent mass of the $i$-th quark (340MeV for $u,d$ and 500MeV for $s$), and $\vec{p_i}$ is the momentum of the $i$-th quark. $V_{conf}$ is the confinement potential. $H_{HF}$ is the hyperfine interaction.

The confinement potential for the pentaquark configurations has been studied in the lattice QCD recently~\cite{Okiharu:2004wy}.
The result shows string-like potential according to the color configurations.
We employ a simpler two-body type potential because five-body potential is much harder to treat.
We use the harmonic oscillator potential:
\begin{equation}
V_{conf} = \sum_{i<j} \frac{1}{2}K | \vec{r}_i-\vec{r}_j |^2, 
\end{equation}
where $K$ is a constant. $\vec{r}_i$ is the coordinate of the $i$-th quark. The harmonic oscillator potential can be used to reproduce the excited baryon spectrum~\cite{Isgur:1978xj,Isgur:1978wd}. The results seem not sensitive to the choice of confinement for the hadron spectrum. We assume that the kinetic term is $SU(3)_F$ symmetric for simplicity. As a result, the orbital wave functions, which are the $1\hbar w$ states of the harmonic oscillator potential, are $SU(3)_{F}$ symmetric. It is known that the $SU(3)_F$ breaking effects from the kinetic term and the wave function are smaller than those from the hyperfine interactions~\cite{Isgur:1978xj}.
The relative $S$-wave state of two-body is described by $\pi^{-3/4}b^{-3/2}\exp\left(-\rho^2/2b^2 \right)$, where $\rho$ is relative coordinate: $\rho = |\vec{r}_1-\vec{r}_2|/\sqrt{2}$ and $b$ is the size parameter: $b = (3Km)^{-1/4}$. 

The hyperfine interactions that we use are given by
\begin{eqnarray}
H_{HF} = (1-P_{III}) H_{OGE}+P_{III}H_{III^{(2)}} + P_{III}H_{III^{(3)}},\label{eq:HF}
\end{eqnarray}
where $H_{OGE}$ is the one-gluon-exchange interaction (OGE)~\cite{DeRujula:1975ge}. $H_{III^{(2)}}$ and $H_{III^{(3)}}$ are the two-body and  three-body terms of III, respectively~\cite{Oka:1989ud}. $P_{III}$ is a parameter which represents the portion of the hyperfine splittings originated from III. The hyperfine splittings come only from  OGE at $P_{III}=0$, while they come entirely from III at $P_{III}=1$.

The spin-spin term of OGE is given by
\begin{eqnarray}
H_{OGE}= \sum_{i<j} V_{ij} \delta^{(3)}(\vec{r}_{ij})  \sigma_i^a \lambda_i^b \sigma_j^a \lambda_j^b,\label{eq:OGE}
\end{eqnarray}
where $V_{ij}$ is the strength of the interaction between the $i$-th quark and $j$-th quark, $\propto (m_i m_j)^{-1}$. 
$\lambda^b_i$ is the color $SU(3)$ Gell-Mann matrix for the $i$-th quark.
For the anti-quark, it means $-\lambda^{b *}_i$. $\sigma^a_i$ is the spin $SU(2)$ Pauli matrix for the $i$-th quark.
Since OGE is the contact interaction, only relative $S$-wave pairs are affected. 
OGE reproduces the baryon and the meson spectra except for the pseudo-scalar mesons.
We determine $V_{ij}$ phenomenologically.
However, it is known that $V_{ij}$ is quite large if $P_{III}=0$ compared with that expected from QCD.
It is favorable that III can reduce the contribution from OGE to the hyperfine interaction.

\begin{figure}
\begin{center}
\includegraphics[width=8.5cm]{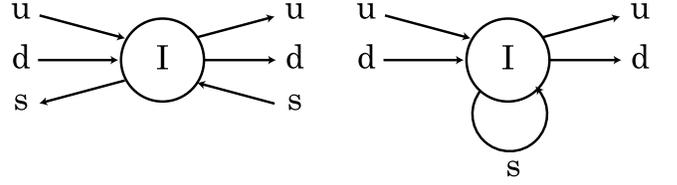}
\end{center}
\caption{\label{fig:III}The three-body interaction among $u,d$ and $\bar{s}$ (left) and the two-body interaction between $u,d$ (right). The latter is obtained by contracting $s\bar{s}$ quark pair from the three-body interaction.}
\end{figure}

We introduce III so that the spectrum of the pseudo-scalar meson also is reproduced. As a result, low-lying hadron spectrum is reproduced. Since the average size of instantons, 1/3 fm, is smaller than that of the hadrons, we assume that III is approximately a contact interaction. 
III contains a three-body interaction and a two-body interaction, which are the determinant type in the flavor space. Thus, the three-body interaction of III affects only the systems of $u$, $d$ and $s$ quarks.
The two-body $qq$ and three-body $qqq$ interactions of III are given~\cite{Oka:1989ud} by
\begin{eqnarray}
H_{III^{(2)}}^{qq} &=& \sum_{i,j=1,i<j}^4 V^{(2)}_{ij} \delta^{(3)}(r_{ij})\left( \mathbf{1}+\frac{3}{32} \lambda_i^b \lambda_j^b \right. \nonumber\\
&+& \left. \frac{9}{32}\sigma_i^a \lambda_i^b \sigma_j^a \lambda_j^b \right), \label{eq:III2}
\end{eqnarray}
\begin{eqnarray}
H_{III^{(3)}}^{qqq}  &=& \sum_{i,j,k=1,i<j<k}^4 V^{(3)}_{ijk} \delta^{(3)}(r_{ij})\delta^{(3)}(r_{ik}) \biggl( \mathbf{1}   \nonumber\\
&+& \frac{3}{32}\left( \lambda_i^b \lambda_j^b + \lambda_j^b \lambda_k^b + \lambda_k^b \lambda_i^b \right) \nonumber\\
&-& \frac{9}{320}  d_{abc} \lambda^a_i \lambda^b_j \lambda^{c}_k \nonumber\\
&+& \frac{9}{32}\left( \sigma_i^a \lambda_i^b \sigma_j^a \lambda_j^b + \sigma_j^a \lambda_j^b \sigma_k^a \lambda_k^b + \sigma_k^a \lambda_k^b \sigma_i^a \lambda_i^b \right) \nonumber\\ 
&+& \frac{27}{320} d_{abc} \lambda^a_i \lambda^b_j\lambda^{c}_k \left( \sigma_i^{\alpha} \sigma_j^{\alpha} + \sigma_j^{\alpha} \sigma_k^{\alpha} + \sigma_k^{\alpha} \sigma_i^{\alpha} \right) \nonumber\\
&-&\left. \frac{9}{64} \epsilon_{\alpha\beta\gamma} f_{abc} \lambda^a_i \lambda^b_j \lambda^{c}_k   \sigma_i^{\alpha} \sigma_j^{\beta} \sigma_k^{\gamma} 
\right).\label{eq:III3}
\end{eqnarray}
where $V^{(2)}_{ij}$ is the strength of the two-body term between the $i$-th quark and $j$-th quark. $V^{(3)}_{ijk}$ is the strength of the three-body term among the $i$-th, $j$-th and $k$-th quarks. $f_{abc}$ and $d_{abc}$ are the $SU(3)$ structure constants defined by $\left[\lambda_a,\lambda_b \right] = 2 i f_{abc} \lambda_c$, $\left\{\lambda_a,\lambda_b \right\} =  4/3 \delta_{ab} I  + 2  d_{abc} \lambda_c$.
The two-body term is obtained by contracting a quark pair from the three-body term as is illustrated in Fig.~\ref{fig:III}.
We obtain the relations between $V^{(2)}$ and $V^{(3)}$ as
\begin{eqnarray}
V^{(2)}_{us} &=& \langle \bar{u}u \rangle V^{(3)}_{uds}, \label{eq:strength3} \\
V^{(2)}_{ud} &=& V^{(2)}_{us}  \frac{\langle \bar{s}s \rangle}{\langle \bar{u}u \rangle} \simeq V^{(2)}_{us}  \frac{m_s}{m_u},
\end{eqnarray}
where the $m_u=m_d$ and $m_s$ are constituent quark masses, $m_u/m_s = 0.6$, and $\langle \bar{u} u \rangle$ and $\langle \bar{s} s \rangle$ are the quark condensates of $u$ and $s$ quarks, respectively: $\langle \bar{u}u \rangle = (-225\mathrm{MeV})^3$. Because the strength of the three-body term is repulsive and the quark condensate is negative, the strength of the two-body term is attractive. Since the three-body term of III is the three-body contact interaction, only relative $S$-wave pairs are affected. Moreover, the determinant type interaction implies that the three-body term of III affects only the flavor singlet states for $3q$.

The interaction between the $\bar{s}$ quark and a $u$ or $d$ quark labeled by $i$ or $j$ can be obtained by the charge conjugation from Eqs.~(\ref{eq:III2}) and (\ref{eq:III3}) :
\begin{eqnarray}
H_{III^{(2)}}^{q\bar{s}} &=& \sum_{i=1}^4 V^{(2)}_{is} \delta^{(3)}(r_{is}) \left( \mathbf{1}-\frac{3}{32}\lambda^b_i(-\lambda_{s}^{b*}) \right. \nonumber \\
&+&\left. \frac{9}{32}\sigma^a_i \lambda^b_{i} \sigma_s^a (-\lambda_{s}^{b*}) \right),\label{eq:III2b}
\end{eqnarray}
\begin{eqnarray}
&&H_{III^{(3)}}^{qq\bar{s}} = \sum_{i,j=1,i<j}^4 V^{(3)}_{ijs} \delta^{(3)}(r_{ij})\delta^{(3)}(r_{js}) \biggl( \mathbf{1}  \nonumber\\
&+& \frac{3}{32}\left( \lambda_i^b \lambda_j^b - \lambda_j^b (-\lambda_s^{b*})-(-\lambda_s^{b*}) \lambda_i^b \right)  \nonumber\\
&+& \frac{9}{320}  d_{abc} \lambda^a_i \lambda^b_j (-\lambda^{*c}_s) \nonumber\\
&+& \frac{9}{32}\left( \sigma_i^a \lambda_i^b \sigma_j^a \lambda_j^b + \sigma_j^a \lambda_j^b \sigma_s^a (-\lambda^{b*}_s) + \sigma_s^a (-\lambda_s^{b*}) \sigma_i^a \lambda_i^b \right)\nonumber\\
&-& \frac{27}{320} d_{abc} \lambda^a_i \lambda^b_j (-\lambda^{*c}_s) \left( \sigma_i^{\alpha} \sigma_j^{\alpha} - \sigma_j^{\alpha} \sigma_s^{\alpha} - \sigma_s^{\alpha} \sigma_i^{\alpha} \right) \nonumber\\
&-&\left. \frac{9}{64} \epsilon_{\alpha\beta\gamma} f_{abc} \lambda^a_i \lambda^b_j (-\lambda^{*c}_s)   \sigma_i^{\alpha} \sigma_j^{\beta} \sigma_s^{\gamma} 
\right).\label{eq:III3b}
\end{eqnarray}

We point out that III contains the three-body term, which is absent in OGE.
The three-body term of III does not change the spectrum of three-quark baryons.
It is, however, expected that III is important in multi-quark systems with strange quarks, such as $2\Lambda$ systems and pentaquarks~\cite{Oka:1989ud,Oka:1990vx,Takeuchi:1990qj,Takeuchi:1992sg,Takeuchi:1993rs,Shinozaki:2004bp}, since they are sensitive to the three-body term of III.
On the other hand, the effects of the  two-body term of III are similar to that of OGE, since the spin dependent forces the last terms of Eqs.~(\ref{eq:III2}) and (\ref{eq:III2b}) are identical to that of OGE~\cite{Shuryak:1988bf,Kochelev:1985de} .
Thus, the baryon spectrum can be reproduced by any combinations of OGE and the two-body term of III.
In contrast, the $\eta-\eta'$ mass splitting is sensitive to III~\cite{Hatsuda:1994pi,Naito:1999sb,Blask:1990ez}.
The $\eta-\eta'$ mass splitting comes from the diagrams of the annihilation type of the two-body term of III, while such diagrams are absent in $\Theta^+$. Therefore, the origin of effects of III for the pentaquark is different from that for $\eta$ and $\eta'$.

Finally, we mention the effects of III for the $S$-wave pentaquarks~\cite{Dmitrasinovic:2005gq}. The anti-symmetrization of $4q$ leads to the unique spin, $S_{4q}=1$. Thus, there are two possible states, a state with $S_{5q}=1/2$ and a state with $S_{5q}=3/2$. Effect of III appears in their mass splitting.
In fact, III reduces the mass splitting as $P_{III}$ increases, given by $225 - 203 P_{III}$ MeV.
Thus, III may favor a possible assignment that $\Theta^+$ has $J^P=3/2^-$\cite{Takeuchi:2004fv,Ishii:2005vc,Nishikawa:2004tk,Lasscock:2005kx}. 

\subsection{P-wave states}
The $J^P=1/2^+$ pentaquarks correspond to the $P$-wave states since the intrinsic parity of the $\bar{s}$ quark is negative.
In order to remove the center-of-mass motion and realize permutation symmetry of the orbital wave function, we use the harmonic oscillator wave function~\cite{Isgur:1978xj,Isgur:1978wd,Stancu:2004du}.
The classification of states is based on the permutation symmetry in the non-relativistic quark model.
$\Theta^+$ consists of four $u,d$ quarks ($4q$) and an $\bar{s}$ quark.
The orbital-spin-isospin-color wave function of $4q$ must be anti-symmetrized.
The wave function of the five quarks ($5q$) must be color singlet and is assumed to be flavor anti-decuplet.

There are two types of $P$-wave $1 \hbar w$ excited states.
In the first type of states~(I), the $\bar{s}$ quark is excited, which is associated with the excitation of the center-of-mass of $4q$ since the center-of-mass of $5q$ must not be excited.
The state~(I) corresponds to the total symmetric states for the orbital wave function of $4q$.
It is known that this state takes a unique spin $S_{4q}=1$ since the total wave function must be the anti-symmetric for $4q$~\cite{Bijker:2003pm}.
The second~(II) type is the one without the $\bar{s}$ quark excited, while an internal coordinate of $4q$ is excited.
The type~(II) corresponds to the [3,1] symmetric states for the orbital wave function of $4q$~\cite{Stancu:2004du}.
The anti-symmetrization for $4q$ leads to no restrictions on $S_{4q}$.

We find that there are nine independent states with $1 \hbar w$ excitation, which consist of two type~(I) states and seven type~(II) states.
We show the nine states in Tables~\ref{tab:base1} and \ref{tab:base2}.
They are further classified into five states with $S_{5q}=1/2$ and four states with $S_{5q}=3/2$. They are classified by the spin-color $SU(6)$ representation~\cite{Jaffe:1976ig,Jaffe:1976ih}.
The first state in Table~\ref{tab:base1} corresponds to the Jaffe-Wilczek (JW) state~\cite{Jaffe:2003sg,Jaffe:2004ph}, which is defined in Appendix~\ref{app:Jaffe-Wilczek Model}. The second row is a state with $S_{4q}=0$, which is in general heavier than the JW state and we refer it to $0^*$. The third row is corresponding to Karliner-Lipkin(KL) state~\cite{Karliner:2003dt,Jennings:2003wz}.
The forth state has $S_{4q}=1$, which we refer to $1^*$. The fifth row is the $\bar{s}$-excited state ( type~(I) ), which we refer to $\bar{s}$.

We only use the spin-spin terms of the hyperfine interactions for simplicity because the $LS$ terms and the tensor terms are known to be weaker than the spin-spin terms.
We point out that the spin-spin terms of the hyperfine interactions do not couple the states with $S_{5q}=1/2$ and the states with $S_{5q}=3/2$.

\begin{table}
\caption{\label{tab:base1}the five states with $S_{5q}=1/2$. $L_{4q}$ is the angular momentum of $4q$ and means that the orbital wave function  $4q$ has a definite permutation symmetry. $S_{4q}$, $S_{5q}$ and $J_{5q}$ are the spin of $4q$ , $5q$ and the total angular momentum of $5q$, respectively. $S_{4q} \times C_{4q}$ is the symmetry of the spin-color $SU(6)$ of $4q$. All states are identified by these quantum numbers.}
\begin{ruledtabular}
\begin{tabular}{c|ccccc}
     & $L_{4q}$&$S_{4q}$&$S_{4q} \times C_{4q}$       & $S_{5q}$      & $J_{5q}$      \\ \hline
JW   &   1 & 0 &\makebox(0,18){}\begin{Young}&&\cr\cr\end{Young}  & $\frac{1}{2}$ & $\frac{1}{2}$ \\
$0^*$&   1 & 0 &\begin{Young}&\cr\cr\cr\end{Young}    & $\frac{1}{2}$ & $\frac{1}{2}$ \\
KL   &   1 & 1 &\begin{Young}&&\cr\cr\end{Young}      & $\frac{1}{2}$ & $\frac{1}{2}$ \\
$1^*$&   1 & 1 &\begin{Young}&\cr\cr\cr\end{Young}    & $\frac{1}{2}$ & $\frac{1}{2}$ \\
$\bar{s}$&0& 1 &\begin{Young}&\cr&\cr\end{Young}      & $\frac{1}{2}$ & $\frac{1}{2}$ \\
\end{tabular}
\end{ruledtabular}
\end{table}

\begin{table}
\caption{\label{tab:base2}the four states with $S_{5q}=3/2$. The spin-spin terms do not couple the states with $S_{5q}=1/2$ and the states with $S_{5q}=3/2$.}
\begin{ruledtabular}
\begin{tabular}{cccccc}
$L_{4q}$&$S_{4q}$&$S_{4q} \times C_{4q}$   & $S_{5q}$      & $J_{5q}$      \\ \hline
1 & 1 & \makebox(0,18){}\begin{Young}&&\cr\cr\end{Young}& $\frac{3}{2}$ & $\frac{1}{2}$ \\
1 & 1 & \begin{Young}&\cr\cr\cr\end{Young} & $\frac{3}{2}$ & $\frac{1}{2}$ \\
1 & 2 & \begin{Young}&\cr\cr\cr\end{Young} & $\frac{3}{2}$ & $\frac{1}{2}$ \\
0 & 1 & \begin{Young}&\cr&\cr\end{Young}   & $\frac{3}{2}$ & $\frac{1}{2}$ \\
\end{tabular}
\end{ruledtabular}
\end{table}

\subsection{Decay}
Decays of the pentaquarks goes through a fall-apart process~\cite{Jaffe:1976ig,Jaffe:1976ih,Jaffe:2004at,Hosaka:2004bn}, which does not require $q\bar{q}$ pair creation.
It is usually expected that the decay widths for the fall-apart decay are much larger than ordinary decays with $q\bar{q}$ creation.

A measure that is often used to estimate the fall-apart width is the $KN$-overlap~\cite{Jennings:2003wz,Carlson:2003xb,Hosaka:2004bn}.
We define the $KN$-overlap by a projection operator for relative $P$-wave $KN$ states,
\begin{eqnarray}
\mathcal{O}_{KN} =  \mathcal{S}_{orb}^{123}\mathcal{S}_{orb}^{4\bar{s}} \mathcal{A}_{color}^{123} \mathcal{A}_{color}^{4\bar{s}} \mathcal{M}_{spin}^{123}\mathcal{A}_{spin}^{4\bar{s}},
\end{eqnarray}
where $\mathcal{S}_{orb}$ is the projection operator to the ground state $K$ or $N$. $\mathcal{A}_{color}$, $\mathcal{M}_{spin}$ and $\mathcal{A}_{spin}$ are the projection operators to color-singlet, the spin 1/2 and the spin 0, respectively. Note that the matrix elements of the operator, $\mathcal{O}_{KN}$, correspond to the absolute square of the $KN$-overlap.

For the four states with $S_{5q}=3/2$, the matrix elements of $\mathcal{O}_{KN}$ are zero since the total spin is different from $KN$. Thus, they can not decay to $KN$ unless a tensor-type interaction is strong.
We obtain the $KN$-overlaps for the five states with $S_{5q}=1/2$ in the bases of Table~\ref{tab:base1},
\begin{eqnarray}
\langle \mathcal{O}_{KN} \rangle = 
\left(
\begin{array}{ccccc}
\frac{5}{192} & \frac{5}{192} & \frac{5\sqrt{3}}{192} & -\frac{5}{192}  & -\frac{\sqrt{10}}{64} \\
\frac{5}{192} & \frac{5}{192} & \frac{5\sqrt{3}}{192} & -\frac{5}{192}  & -\frac{\sqrt{10}}{64} \\
\frac{5\sqrt{3}}{192} & \frac{5\sqrt{3}}{192} & \frac{5}{64} & -\frac{5\sqrt{3}}{192} & -\frac{\sqrt{30}}{64} \\
-\frac{5}{192}& -\frac{5}{192} & -\frac{5\sqrt{3}}{192} & \frac{5}{192} & \frac{\sqrt{10}}{64} \\
 -\frac{\sqrt{10}}{64} & -\frac{\sqrt{10}}{64} & -\frac{\sqrt{30}}{64} & \frac{\sqrt{10}}{64} & \frac{3}{32} \\
\end{array}
\right). \label{eq:OKN}
\end{eqnarray}
We diagonalize this matrix with a unitary matrix $U$:
\begin{eqnarray}
\langle \mathcal{O}_{KN} \rangle &=& U^{-1}
\left(
\begin{array}{ccccc}
\frac{1}{4}&0&0&0&0\\
0&0&0&0&0\\
0&0&0&0&0\\
0&0&0&0&0\\
0&0&0&0&0\\
\end{array}
\right)U.
\end{eqnarray}
We find that a state, which we call ``$KN$ state'', has non-zero $KN$-overlap, while the other four states, called ``confined states'', have no $KN$-overlaps. The ``confined states'' can not directly fall apart to $KN$,
while  the ``$KN$ state'' is expected to couple strongly to the $KN$ scattering state.
( A model conclusion of the decay width according to the fall-apart process is achieved by using the meson-quark-quark vertex~\cite{Hosaka:2004bn}.) 
The ``$KN$ state'' is given by an eigen-vector of Eq.~(\ref{eq:OKN}):
\begin{eqnarray}
| KN \rangle = \left(
\begin{array}{ccccc}
\frac{\sqrt{15}}{12},&\frac{\sqrt{15}}{12},&\frac{3\sqrt{5}}{12},&-\frac{\sqrt{15}}{12},&-\frac{3\sqrt{6}}{12}
\end{array}
\right).
\end{eqnarray}
This state corresponds to the anti-symmetrized $K$ and $N$ with $1 \hbar w$ excitation : $| KN \rangle = \mathcal{A}^{1234}\left( N_{123} \otimes K_{4\bar{s}} \right)$, where $\mathcal{A}^{1234}$ is anti-symmetrizer of $4q$.

The couplings between the ``$KN$ state'' and the ``confined states'' can be described by channel coupling scattering formalism~\cite{Karliner:2004qw,Jaffe:2004at}.
We assume that the pentaquark states are given only by the ``confined states'', $i.e.$ the pentaquark states are eigenstates of the Hamiltonian within the subspace of the ``confined states''.
We have to diagonalize only the hyperfine interaction in the subspace of the ``confined states'', resulting in 
\begin{eqnarray}
&&\langle H_{HF} \rangle = \nonumber\\
&&\left(
\begin{array}{ccccc}
E_{KN}& a_{KN\Theta^+} &a_{KN\Theta^*_1}&a_{KN\Theta^*_2}&a_{KN\Theta^*_3}\\
a_{KN\Theta^+}  &E_{\Theta^+}&0&0&0\\
a_{KN\Theta^*_1}&0&E_{\Theta^*_1}&0&0\\
a_{KN\Theta^*_2}&0&0&E_{\Theta^*_2}&0\\
a_{KN\Theta^*_3}&0&0&0&E_{\Theta^*_3}\\
\end{array}\right),
\end{eqnarray}
where all the values are functions of $P_{III}$. $E_{\Theta}$'s are the eigenvalues after the diagonalization in the ``confined space''.
The off-diagonal elements, $a_{KN\Theta}$, correspond to couplings between the ``$KN$ state'' and ``confined states''.
Note that in the present model, no other terms in the Hamiltonian have off-diagonal matrix elements. 
We assign the lowest state in the ``confined states'' to $\Theta^+$.
Then, the narrow width of the pentaquarks can be attributed to a small $a_{KN\Theta^+}$.
The other seven states, $\Theta^*_1$, $\Theta^*_2$, $\Theta^*_3$ and the four states with $S_{5q}=3/2$, are regarded as excited states of $\Theta^+$. All the states have $J^P=1/2^+$.

\section{\label{sec:Results}Results}
We investigate spectrum of the $P$-wave pentaquarks with $I=0$ and $J^P=1/2^+$ in the non-relativistic quark model with instanton induced interaction (III).
We assume that the kinetic term and the orbital wave functions are $SU(3)_F$ symmetric for simplicity. Furthermore, we neglect the $LS$ terms and the tensor terms.

Then, the pentaquarks are obtained by diagonalizing the hyperfine interaction in the subspace of the ``confined states''.
Both the masses and the couplings are obtained simultaneously.
Because the five-quark confinement may allow an extra constant, we have discuss only the excitation energies and the structures of the obtained states.
Thus, we set the mass of $\Theta^+$ to 1540MeV.
It should be noted that if we use the confinement potential derived from the baryon spectrum without adjusting a constant term, then the absolute mass of $\Theta^+$ in this model is about 2GeV.
$P_{III}$ can be determined from the $\eta-\eta'$ mass splitting, which gives $P_{III}=0.3-0.5$~\cite{Takeuchi:1992sg}.
However, the non-relativistic quark model gives large ambiguities for the pseudo-scalar mesons.
Thus, we here treat $P_{III}$ as a free parameter.
The size parameter, $b$, is unknown for the pentaquarks.
We use $b \approx 0.5$fm, which is taken from that of the nucleon.
It is known that the radii of the $S$-wave pentaquarks are as small as that of the nucleon if III is introduced~\cite{Shinozaki:2004bp}.
The strengths of OGE and the two-body term of III are fixed phenomenologically from the $N-\Delta$ mass splitting so that they reproduce the baryon and meson spectra except for the pseudo-scalar mesons.
The strength of the three-body term of III is determined by Eq.~(\ref{eq:strength3}).
They give
\begin{eqnarray}
&&\langle V_{ud} \rangle = -19\mbox{MeV},\ \  \langle V_{us} \rangle = -11\mbox{MeV}, \nonumber \\
&&\langle V_{ud}^{(2)} \rangle = -67\mbox{MeV},\ \langle V_{us}^{(2)}\rangle = -40\mbox{MeV},\nonumber\\
&&\langle V_{uds}^{(3)} \rangle = + 20 \mbox{MeV}.
\end{eqnarray}

The matrix elements of the hyperfine interaction, $H_{HF}$, for the nine states in Tables~\ref{tab:base1} and \ref{tab:base2} are given in Appendix~\ref{app:The matrix elements}.
Since the other terms of the Hamiltonian, Eq.~(\ref{eq:Hamiltonian}), are identical for the nine states,
the spectrum is determined only by the hyperfine interaction.
We find that both OGE and the two-body term of III play similar roles in the spectrum.
A small difference comes from the first and second terms in Eqs.~(\ref{eq:III2}) and (\ref{eq:III2b}).
Therefore, the hyperfine interaction of Eq.~(\ref{eq:HF}) is approximately described as $H_{HF} = H^{(2)} + P_{III}H_{III^{(3)}}$, where $H^{(2)}$ contains both OGE and the two-body term of III and is approximately independent of $P_{III}$.
Thus, the effects of III are attributed to the three-body term.
We find that the contribution of the three-body term of III is repulsive and smaller than those of $H^{(2)}$ for the pentaquarks.

\subsection{Spectrum}

\begin{figure}
\begin{center}
\includegraphics[width=0.48\textwidth]{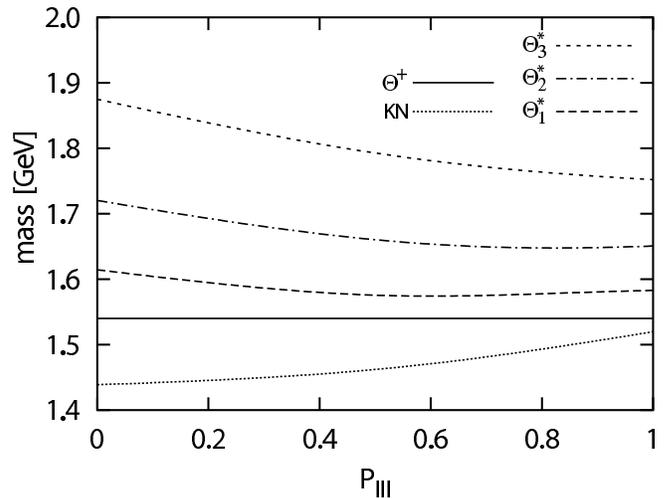}
\caption{\label{fig:hyperfine1}The masses of the pentaquarks with $S_{5q}=1/2$. The mass of $\Theta^+$ is set to 1540MeV. $\Theta^*_1$, $\Theta^*_2$ and $\Theta^*_3$ are excited states of $\Theta^+$. $KN$ is the ``$KN$ state'', which is very broad. }
\end{center}
\end{figure}

We show the masses of pentaquarks with $S_{5q}=1/2$, in Fig.~\ref{fig:hyperfine1}. 
Since the three-body term of III is weakly repulsive, the hyperfine splittings are reduced as $P_{III}$ increases.
We set the mass of $\Theta^+$ to 1540MeV.
Note that the ``$KN$ state'' is always above the $KN$ threshold.
Thus, the ``$KN$ state'' may not form a resonance.  The excited states of $\Theta^+$ lie at about 1.6GeV, 1.7GeV and 1.8GeV. If the widths are narrow, those states should be observed. 

\begin{figure}
\begin{center}
\includegraphics[width=0.48\textwidth]{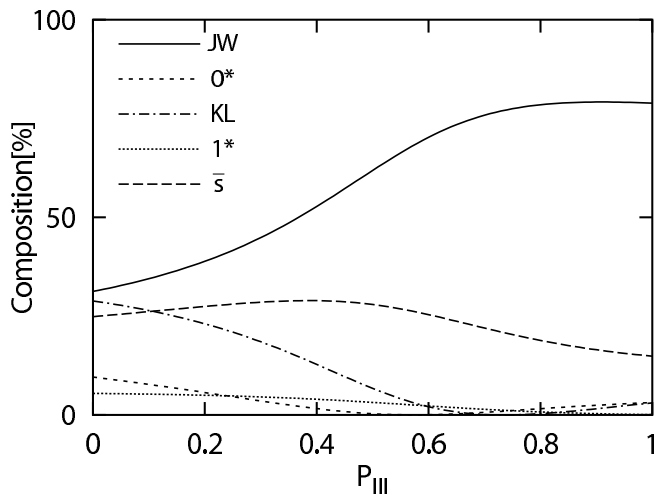}
\caption{\label{fig:composition}The composition of $\Theta^+$ in the bases of Table~\ref{tab:base1}.
The solid line is the JW state. The JW state is dominant for large $P_{III}$, while the KL state is suppressed. }
\end{center}
\end{figure}

Fig.~\ref{fig:composition} shows the decomposition of the lowest $\Theta^+$ state in the bases of Table~\ref{tab:base1}. 
We find that the JW state is dominant at large $P_{III}$, while the KL state is suppressed.
The contributions from the heavier states, $0^*$ and $1^*$, are very small.
The $P_{III}$ dependence can be understood as follows.
OGE and the two-body term of III are more attractive for both the JW state and the KL state. However, the three-body term of III for the JW state is less repulsive than that for the KL state.
The tri-quark of the KL state is strongly affected by the three-body term of III, since the $u, d$ and $\bar{s}$ quarks are in relative $S$-wave states.
In contrast, the JW state is less sensitive since the $\bar{s}$ quark is separated from the $u,d$ quarks.
At large $P_{III}$, $\Theta^+$ attains nearly the maximum JW component.
Thus, we find that the JW state is a favorable state with respect to III.

\subsection{Decay}
\begin{figure}
\begin{center}
\includegraphics[width=0.48\textwidth]{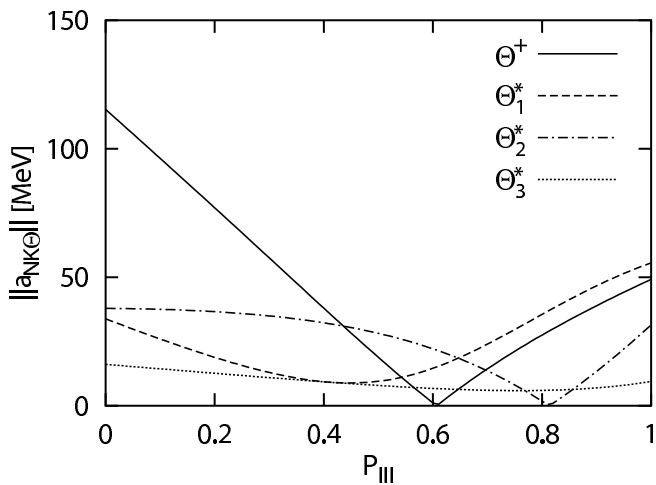}
\caption{\label{fig:coupling}The couplings between the ``$KN$ state'' and the pentaquarks with $S_{5q}=1/2$.
The coupling of $\Theta^+$ is zero at $P_{III} = 0.61$. At the zero point, $\Theta^+$ can not decay to $KN$. The $\Theta^*_2$ is stable at $P_{III}=0.81$.}
\end{center}
\end{figure}

Fig.~\ref{fig:coupling} shows the couplings between the ``$KN$ state'' and the pentaquarks with $S_{5q}=1/2$.
The coupling of $\Theta^+$ is very large in the case without III, $P_{III}=0$, while the coupling becomes much weaker as $P_{III}$ increases.
In the relevant range of $P_{III}$, the coupling is about 1/3 of that at  $P_{III}=0$.
It hits zero at $P_{III} = 0.61$, where $\Theta^+$ does not couple to the ``$KN$ state'', that is, $\Theta^+$ becomes stable against the decay to $KN$ in the present model.
At $P_{III} = 0.81$, the second resonance $\Theta^*_2$ becomes stable. The other states, $\Theta^*_1$ and $\Theta^*_3$, do not become stable within $0\le P_{III} \le1$.
We point out that the couplings are non-zero in most range of $P_{III}$, and therefore no more than one pentaquark becomes simultaneously stable.
This may explain why only one pentaquarks has been seen.

\subsection{The states with $S_{5q}=3/2$}

\begin{figure}
\begin{center}
\includegraphics[width=0.48\textwidth]{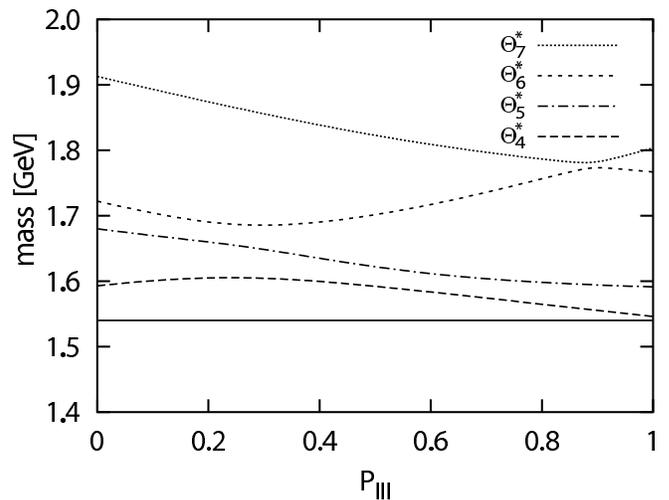}
\caption{\label{fig:hyperfine2}The masses of the pentaquarks with $S_{5q}=3/2$, which are excited states of $\Theta^+$.
The solid line is $\Theta^+$, which is 1540MeV. The pentaquarks lie above $\Theta^+$. }
\end{center}
\end{figure}

The spin-spin terms of the hyperfine interactions do not change the total spin and therefore do not mix $S_{5q}=1/2$ and $S_{5q}=3/2$.
In Fig.~\ref{fig:hyperfine2}, the masses of the $S_{5q}=3/2$ states are plotted.
They lie at about 1.6GeV, 1.65GeV, 1.7GeV and above 1.8GeV.
OGE and the two-body term of III for the pentaquarks with $S_{5q}=1/2$ are more attractive than those with $S_{5q}=3/2$.
Thus, the lowest state, $\Theta^+$, has always $S_{5q}=1/2$.
It is also seen that the effects of the three-body term of III are strong in $S_{5q}=3/2$.
One sees several level-crossings when $P_{III}$ goes from 0 to 1.
The state which rises with $P_{III}$  corresponds to the most repulsive eigen-state of the three-body term of III.
Since the spin is different from the ``$KN$ states'', the $S_{5q}=3/2$ states decay to $KN$ only through a tensor force, which is expected to be weak in the present quark model.

\section{\label{sec:Conclusion}Conclusion}
We have investigated the spectrum of the $P$-wave pentaquarks with $I=0$ and $J^P=1/2^+$ in the non-relativistic quark model with instanton induced interaction (III).
We expect that effects of the two-body term of III should appear in the $I=0$ meson channels,
while the effects of the three-body III may be important in multi-quark systems with strange quarks.

We find that there are nine $1\hbar w$ states in the harmonic oscillator potential.
We find that one of them, the ``$KN$ state'', has a finite overlap with $KN$, while eight other  ``confined states'' have no $KN$-overlaps.
The pentaquarks can decay to $KN$ only through the ``$KN$ state''.
We assign the lowest energy eigen-state in the ``confined states'' to $\Theta^+$.
We point out that the decay is not only a fall-apart process, and introduce channel coupling scattering between the ``$KN$ state'' and the ``confined states'',
where the couplings are determined only by the hyperfine interaction under our assumptions.
Since the fall-apart leads to the large widths, the narrow widths of the pentaquarks are explained by small couplings of the channel coupling scattering.

We find that III reduces the coupling between $\Theta^+$ and the ``$KN$ state''.
As the contributions of III become larger, $\Theta^+$ becomes significantly stable and the dominant component of $\Theta^+$ becomes the JW state.
We therefore give a qualitative explanation of the narrow width of $\Theta^+$ and we find the dynamics that the JW state becomes naturally the dominant component.
III changes the composition of $\Theta^+$.
In other words, III leads to a cancellation of the coupling between $\Theta^+$ and $KN$, where OGE, the two-body term of III and the three-body term of III are canceled out each other.

We find that at $P_{III}=0.61$, $\Theta^+$ becomes stable within our approach.
This value of $P_{III}$ is close to the relevant range, $P_{III}=0.3-0.5$, which is consistent with the $\eta-\eta'$ mass splitting. 

We point out that the effects of III strongly depend on the flavor part of the system.
For example a pentaquark with $I=2$, $uuuu\bar{s}$, is not affected by the three-body term of III.
We conjecture that the flavor dependence is a reason why no stable pentaquark is seen in other channels. 
In order to make the predictions more realistic, the $SU(3)_F$ breaking effects of the kinetic term and the orbital wave function must be taken into account.
The other pentaquarks in the $\bar{10}_F$ representation, $N_{\bar{10}},\Sigma_{\bar{10}}$ and $\Phi$, may be affected by the $SU(3)_F$ breaking effects. 
Again it is probable that those pentaquarks do not become stable simultaneously, which may explain why the other members of the $\bar{10}_F$ representation is not observed.
It is also noted that the mixing between $\bar{10}_F$ and $8_F$ representations is also sensitive to the two-body term of III,
since the mixing matrix elements include flavor singlet $q\bar{q}$ contributions.

Further effects of III will appear in diquark correlations in the pentaquarks, which are not included in the present harmonic oscillator wave functions.
We expect that III makes the diquarks compact. The three-body term of III gives a repulsion between a diquark and the $\bar{s}$ quark. Thus, we expect that the $KN$-overlap is reduced and the JW state is more favorable.

The tensor terms of OGE and III must be taken into account in order to evaluate the mixings of $S_{5q}=1/2$ and $S_{5q}=3/2$. 
Such mixing is favorable for a small width since the $S_{5q}=3/2$ states does not couple directly to $KN$.
Predictions and searches of the $LS$ partner of $\Theta^+$ are important.
If the $LS$ terms for the pentaquarks are significantly weak similarly to that for the $P$-wave baryons, the $LS$ partner must be observed.

Finally, we conclude that III affects significantly not only $\eta$ and $\eta'$, but also pentaquarks.
Further studies of the pentaquarks should help a deeper understanding of dynamics of QCD.

\begin{acknowledgments}
We would like to thank A.Hosaka, K.Shimizu, N.Ishii and V.Dmitrasinovic for useful discussions.
M.~O. acknowledges supports from the Grants-in-Aid for Scientific Research in Priority Areas 1707002, and (B) 15340072.
S.~T is supported by the Grants-in-Aid (c) 15540289.
T.~S. is supported by the 21st Century COE Program at Tokyo Tech ``Nanometer-Scale Quantum Physics" by the Ministry of Education, Culture, Sports, Science and Technology.
\end{acknowledgments}

\appendix

\section{\label{app:Jaffe-Wilczek Model}Jaffe-Wilczek Model}
The Jaffe and Wilczek (JW)~\cite{Jaffe:2003sg} proposed a picture of the pentaquark, a resonance of two scalar-isoscalar diquarks and an $\bar{s}$ quark. This model has a unique symmetry structure. 

The diquark consists of $u$ and $d$ quarks ($2q$), which form a relative $S$-wave pair with the spin $S_{2q}=0$, the isospin $I_{2q}=0$ and the color $C_{2q}=\bar{3}$.
Possible symmetries in the spin$\times$color space of $4q$ are
\begin{equation}
\begin{array}{ccccccc}
\begin{Young}&\cr&\cr\end{Young} &\times& \begin{Young}&\cr\cr\cr\end{Young} &=&
\begin{Young}&&\cr\cr\end{Young} &+& \begin{Young}&\cr\cr\cr\end{Young} \\
S_{4q}=0  & & C_{4q}=3& & & &
\end{array},
\end{equation}
while possible symmetries in the spin$\times$color $(SC)$ space of the two diquarks are
\begin{equation}
\begin{array}{ccccccccc}
\begin{Young}&\cr\end{Young}  & \otimes & \begin{Young}&\cr\end{Young} &=&
\begin{Young}&&&\cr\end{Young} &+& \begin{Young}&&\cr\cr\end{Young} &+& \begin{Young}&\cr&\cr\end{Young} \\
SC_{2q} && SC_{2q} &  &
\end{array}.
\end{equation}
Thus the symmetry of spin$\times$color of $4q$ is uniquely determined :
\begin{equation}
\begin{array}{ccccc}
\begin{Young}&\cr&\cr\end{Young} &\times& \begin{Young}&\cr\cr\cr\end{Young} &\to&
\begin{Young}&&\cr\cr\end{Young} \\
S_{4q}=0  & & C_{4q}=3& & 
\end{array}.
\end{equation}
When the isospin symmetry is combined, we have spin$\times$color$\times$isospin symmetries of $4q$ :
\begin{equation}
\begin{array}{ccccccc}
\begin{Young}&&\cr\cr\end{Young}  & \times & \begin{Young}&\cr&\cr\end{Young} &=&
\begin{Young}&&\cr\cr\end{Young} &+& \begin{Young}&\cr\cr\cr\end{Young} \\
SC_{4q} && I_{4q}=0 & & & & 
\end{array}.
\end{equation}
On the other hand, possible symmetries of the orbital wave function $(O)$ of the two diquarks are
\begin{equation}
\begin{array}{ccccccccc}
\begin{Young}&\cr\end{Young} &\otimes& \begin{Young}&\cr\end{Young} &=&
\begin{Young}&&&\cr\end{Young} &+& \begin{Young}&&\cr\cr\end{Young} &+& \begin{Young}&\cr&\cr\end{Young} \\
O_{2q} && O_{2q} &  & 
\end{array}.
\end{equation}
Since the symmetry of the orbital wave function must be the conjugate to that of the spin$\times$color$\times$isospin, 
it is uniquely determined :
\begin{equation}
\begin{array}{ccccc}
\begin{Young}&\cr\end{Young} &\otimes& \begin{Young}&\cr\end{Young} &\to&
 \begin{Young}&&\cr\cr\end{Young} \\
O_{2q} && O_{2q} &  & 
\end{array}.
\end{equation}

Finally, we obtain the symmetry of the JW wave function:
\begin{equation}
\begin{array}{c @{\hspace{10pt}} c @{\hspace{10pt}} c @{\hspace{10pt}} c @{\hspace{10pt}} c}
\begin{Young}&&\cr\cr\end{Young} & \begin{Young}&\cr&\cr\end{Young} & \begin{Young}&\cr&\cr\end{Young} & \begin{Young}&\cr\cr\cr\end{Young}&\begin{Young}&&\cr\cr\end{Young}\\
O_{4q} & S_{4q} & I_{4q} & C_{4q} & SC_{4q}
\end{array}.
\end{equation}

It is noted that in the spin$\times$flavor ($SI_{4q}$) space of $4q$, the JW state does not belong to any definite symmetry. 

\section{\label{app:The matrix elements}The matrix elements}
We show the matrix elements of OGE and III for the states with $S_{5q}=1/2$ (Table~\ref{tab:base1}) and the states with $S_{5q}=3/2$ (Table~\ref{tab:base2}). The matrix elements depend on the orbital wave functions.
We assume that the orbital wave functions are the $1\hbar w$ states of the harmonic oscillator potential and $SU(3)_{F}$ symmetric.

We obtain the matrix elements of OGE in Eq.~(\ref{eq:OGE}):
\begin{eqnarray}
&&\langle H_{OGE}^{S_{5q}=1/2} \rangle = \langle V_{ud} \rangle \times \nonumber \\
&&\left(
\begin{array}{ccccc}
13   & 3 & \frac{20}{\sqrt{3}}\xi & \xi  &\frac{-15}{\sqrt{10}}\xi \\
3 & \frac{7}{3} & \frac{-4}{\sqrt{3}}\xi & \frac{-19}{3}\xi & \frac{-35}{3\sqrt{10}}\xi \\
\frac{20}{\sqrt{3}}\xi  & \frac{-4}{\sqrt{3}}\xi & \frac{46+63\xi}{6} &  \frac{-10+5 \xi}{2\sqrt{3}} & \frac{-25}{\sqrt{30}}\xi \\
\xi & \frac{-19}{3}\xi & \frac{-10+5 \xi}{2\sqrt{3}} & \frac{-6-19\xi}{6} & \frac{7\sqrt{10}}{6} \xi \\
\frac{-15}{\sqrt{10}}\xi & \frac{-35}{3\sqrt{10}}\xi & \frac{-25}{\sqrt{30}}\xi & \frac{7\sqrt{10}}{6} \xi & \frac{16+15\xi}{3} \\
\end{array}\right),
\end{eqnarray}
\begin{eqnarray}
&&\langle H_{OGE}^{S_{5q}=3/2} \rangle = \langle V_{ud} \rangle \times \nonumber\\
&&\left(
\begin{array}{cccc}
\frac{92-63\xi}{12} & \frac{-20-5\xi}{4\sqrt{3}} & \frac{\sqrt{30}}{12}\xi & \frac{5\sqrt{30}}{12}\xi\\
\frac{-20-5\xi}{4\sqrt{3}} & \frac{-12+19\xi}{12} & \frac{-47\sqrt{10}}{12}\xi & \frac{-7\sqrt{10}}{12}\xi \\
\frac{\sqrt{30}}{12}\xi & \frac{-47\sqrt{10}}{12}\xi & \frac{-28+34\xi}{6}  &-\frac{5}{6}\xi \\
\frac{5\sqrt{30}}{12}\xi & \frac{-7\sqrt{10}}{12}\xi &-\frac{5}{6}\xi & \frac{32-15\xi}{6} \\
\end{array}\right),
\end{eqnarray}
where $\xi=\langle V_{us} \rangle /\langle V_{ud} \rangle  = m_u/m_s$. $\langle V_{ud} \rangle = \langle V_{ij} \delta^{(3)}(\vec{r}_{ij}) \rangle_{orb} = V_{ud}/\sqrt{2 \pi b^2}^3$ contains the spatial integration between $u$ and $d$ quarks, which depends on the sizes of the pentaquarks. $\langle V_{us} \rangle $ is similarly given.

We obtain the matrix elements of the two-body term of III in Eqs.~(\ref{eq:III2}) and (\ref{eq:III2b}):
\begin{eqnarray}
&&\langle H_{III^{(2)}}^{S_{5q}=1/2} \rangle = \langle V^{(2)}_{ud} \rangle \times \nonumber\\
&&\left(
\begin{array}{ccccc}
\frac{444+249\xi}{64} & \frac{36-9\xi}{64}  & \frac{15\sqrt{3}}{8}\xi & \frac{9}{32}\xi & \frac{-135}{32\sqrt{10}}\xi\\
\frac{36-9\xi}{64}  & \frac{252+249\xi}{64} & -\frac{3\sqrt{3}}{8}\xi & -\frac{57}{32}\xi & \frac{-105}{32\sqrt{10}}\xi\\
\frac{15\sqrt{3}}{8}\xi  &  -\frac{3\sqrt{3}}{8}\xi  & \frac{180+219\xi}{32} & \frac{-36+27\xi}{32\sqrt{3}} &\frac{-135}{16\sqrt{30}}\xi\\
\frac{9}{32}\xi & - \frac{57}{32}\xi &  \frac{-36+27\xi}{32\sqrt{3}}  & \frac{108+99\xi}{32}  &\frac{15}{4\sqrt{10}}\xi\\
\frac{-135}{32\sqrt{10}}\xi&\frac{-105}{32\sqrt{10}}\xi&\frac{-135}{16\sqrt{30}}\xi&\frac{15}{4\sqrt{10}}\xi&\frac{216+99\xi}{32} \\
\end{array}\right),
\end{eqnarray}
\begin{eqnarray}
&&\langle  H_{III^{(2)}}^{S_{5q}=3/2} \rangle = \langle  V^{(2)}_{ud} \rangle \times\nonumber\\
&&\left(
\begin{array}{cccc}
\frac{720+309\xi}{128} &  \frac{-144-27\xi}{128\sqrt{3}}  & \frac{3\sqrt{30}}{128}\xi  & \frac{9\sqrt{30}}{128}\xi \\
\frac{-144-27\xi}{128\sqrt{3}}  & \frac{432+567\xi}{128} &  \frac{-141\sqrt{10}}{128} \xi & \frac{-15\sqrt{10}}{128} \xi \\
\frac{3\sqrt{30}}{128}\xi & \frac{-141\sqrt{10}}{128} \xi & \frac{144+411\xi}{64} &-\frac{15}{64}\xi\\
\frac{9\sqrt{30}}{128}\xi & \frac{-15\sqrt{10}}{128} \xi &-\frac{15}{64}\xi& \frac{432+63\xi}{64}  \\
\end{array}\right).
\end{eqnarray}

We obtain the matrix elements of the three-body term of III in Eq.~(\ref{eq:III3b}):
\begin{eqnarray}
&&\langle  H_{III^{(3)}}^{S_{5q}=1/2} \rangle  = \langle  V^{(3)}_{uds} \rangle \times \nonumber\\
&&\left(
\begin{array}{ccccc}
\frac{69}{8}   & 0 & \frac{39\sqrt{3}}{8}  & 0 & \frac{-45}{4\sqrt{10}}  \\
0 & \frac{27}{8}   & \frac{-27}{8\sqrt{3}}  & 0 &0\\
\frac{39\sqrt{3}}{8}   & \frac{-27}{8\sqrt{3}}    & 12  & 0 & \frac{-45}{4\sqrt{30}} \\
0 & 0 & 0 & 0 & 0\\
\frac{-45}{4\sqrt{10}} & 0 & \frac{-45}{4\sqrt{30}}  & 0 & \frac{9}{4}  
\end{array}\right),
\end{eqnarray}
\begin{eqnarray}
&&\langle H_{III^{(3)}}^{S_{5q}=3/2} \rangle = \langle  V^{(3)}_{uds} \rangle \times \nonumber\\
&&\left(
\begin{array}{cccc}
\frac{69}{32}   & \frac{-27\sqrt{3}}{32} & \frac{9\sqrt{30}}{32} & \frac{3\sqrt{30}}{16} \\
\frac{-27\sqrt{3}}{32} & \frac{243}{32}  &  \frac{-81\sqrt{10}}{32} &0\\
\frac{9\sqrt{30}}{32}  & \frac{-81\sqrt{10}}{32}  & \frac{135}{16} &0\\
\frac{3\sqrt{30}}{16}  & 0 & 0 & \frac{9}{16}  \\
\end{array}\right),
\end{eqnarray}
where $\langle V_{uds} \rangle = \langle V_{ijs}^{(3)} \delta^{(3)}(\vec{r}_{ij}) \delta^{(3)}(\vec{r}_{js}) \rangle_{orb} = V_{uds}/\sqrt{3\pi^2 b^4}^3$ contains the spatial integration among $u,d$ quarks and $\bar{s}$ quark.

\bibliography{shinozk}

\begin{thebibliography}{58}
\expandafter\ifx\csname natexlab\endcsname\relax\def\natexlab#1{#1}\fi
\expandafter\ifx\csname bibnamefont\endcsname\relax
  \def\bibnamefont#1{#1}\fi
\expandafter\ifx\csname bibfnamefont\endcsname\relax
  \def\bibfnamefont#1{#1}\fi
\expandafter\ifx\csname citenamefont\endcsname\relax
  \def\citenamefont#1{#1}\fi
\expandafter\ifx\csname url\endcsname\relax
  \def\url#1{\texttt{#1}}\fi
\expandafter\ifx\csname urlprefix\endcsname\relax\def\urlprefix{URL }\fi
\providecommand{\bibinfo}[2]{#2}
\providecommand{\eprint}[2][]{\url{#2}}

\bibitem[{\citenamefont{Nakano et~al.}(2003)}]{Nakano:2003qx}
\bibinfo{author}{\bibfnamefont{T.}~\bibnamefont{Nakano}} \bibnamefont{et~al.}
  (\bibinfo{collaboration}{LEPS}), \bibinfo{journal}{Phys. Rev. Lett.}
  \textbf{\bibinfo{volume}{91}}, \bibinfo{pages}{012002}
  (\bibinfo{year}{2003}), \eprint{hep-ex/0301020}.

\bibitem[{\citenamefont{Danilov}(2005)}]{Danilov:2005kt}
\bibinfo{author}{\bibfnamefont{M.}~\bibnamefont{Danilov}}
  (\bibinfo{year}{2005}), \eprint{hep-ex/0509012}.

\bibitem[{\citenamefont{Diakonov et~al.}(1997)\citenamefont{Diakonov, Petrov,
  and Polyakov}}]{Diakonov:1997mm}
\bibinfo{author}{\bibfnamefont{D.}~\bibnamefont{Diakonov}},
  \bibinfo{author}{\bibfnamefont{V.}~\bibnamefont{Petrov}}, \bibnamefont{and}
  \bibinfo{author}{\bibfnamefont{M.~V.} \bibnamefont{Polyakov}},
  \bibinfo{journal}{Z. Phys.} \textbf{\bibinfo{volume}{A359}},
  \bibinfo{pages}{305} (\bibinfo{year}{1997}), \eprint{hep-ph/9703373}.

\bibitem[{\citenamefont{Jaffe and Wilczek}(2003)}]{Jaffe:2003sg}
\bibinfo{author}{\bibfnamefont{R.~L.} \bibnamefont{Jaffe}} \bibnamefont{and}
  \bibinfo{author}{\bibfnamefont{F.}~\bibnamefont{Wilczek}},
  \bibinfo{journal}{Phys. Rev. Lett.} \textbf{\bibinfo{volume}{91}},
  \bibinfo{pages}{232003} (\bibinfo{year}{2003}), \eprint{hep-ph/0307341}.

\bibitem[{\citenamefont{Jennings and Maltman}(2004)}]{Jennings:2003wz}
\bibinfo{author}{\bibfnamefont{B.~K.} \bibnamefont{Jennings}} \bibnamefont{and}
  \bibinfo{author}{\bibfnamefont{K.}~\bibnamefont{Maltman}},
  \bibinfo{journal}{Phys. Rev.} \textbf{\bibinfo{volume}{D69}},
  \bibinfo{pages}{094020} (\bibinfo{year}{2004}), \eprint{hep-ph/0308286}.

\bibitem[{\citenamefont{Bijker et~al.}(2004)\citenamefont{Bijker, Giannini, and
  Santopinto}}]{Bijker:2003pm}
\bibinfo{author}{\bibfnamefont{R.}~\bibnamefont{Bijker}},
  \bibinfo{author}{\bibfnamefont{M.~M.} \bibnamefont{Giannini}},
  \bibnamefont{and}
  \bibinfo{author}{\bibfnamefont{E.}~\bibnamefont{Santopinto}},
  \bibinfo{journal}{Eur. Phys. J.} \textbf{\bibinfo{volume}{A22}},
  \bibinfo{pages}{319} (\bibinfo{year}{2004}), \eprint{hep-ph/0310281}.

\bibitem[{\citenamefont{Carlson et~al.}(2003)\citenamefont{Carlson, Carone,
  Kwee, and Nazaryan}}]{Carlson:2003pn}
\bibinfo{author}{\bibfnamefont{C.~E.} \bibnamefont{Carlson}},
  \bibinfo{author}{\bibfnamefont{C.~D.} \bibnamefont{Carone}},
  \bibinfo{author}{\bibfnamefont{H.~J.} \bibnamefont{Kwee}}, \bibnamefont{and}
  \bibinfo{author}{\bibfnamefont{V.}~\bibnamefont{Nazaryan}},
  \bibinfo{journal}{Phys. Lett.} \textbf{\bibinfo{volume}{B573}},
  \bibinfo{pages}{101} (\bibinfo{year}{2003}), \eprint{hep-ph/0307396}.

\bibitem[{\citenamefont{Karliner and Lipkin}(2003)}]{Karliner:2003dt}
\bibinfo{author}{\bibfnamefont{M.}~\bibnamefont{Karliner}} \bibnamefont{and}
  \bibinfo{author}{\bibfnamefont{H.~J.} \bibnamefont{Lipkin}},
  \bibinfo{journal}{Phys. Lett.} \textbf{\bibinfo{volume}{B575}},
  \bibinfo{pages}{249} (\bibinfo{year}{2003}), \eprint{hep-ph/0402260}.

\bibitem[{\citenamefont{Kochelev et~al.}(2004)\citenamefont{Kochelev, Lee, and
  Vento}}]{Kochelev:2004nd}
\bibinfo{author}{\bibfnamefont{N.~I.} \bibnamefont{Kochelev}},
  \bibinfo{author}{\bibfnamefont{H.~J.} \bibnamefont{Lee}}, \bibnamefont{and}
  \bibinfo{author}{\bibfnamefont{V.}~\bibnamefont{Vento}},
  \bibinfo{journal}{Phys. Lett.} \textbf{\bibinfo{volume}{B594}},
  \bibinfo{pages}{87} (\bibinfo{year}{2004}), \eprint{hep-ph/0404065}.

\bibitem[{\citenamefont{Kanada-Enyo et~al.}(2005)\citenamefont{Kanada-Enyo,
  Morimatsu, and Nishikawa}}]{Kanada-Enyo:2004bn}
\bibinfo{author}{\bibfnamefont{Y.}~\bibnamefont{Kanada-Enyo}},
  \bibinfo{author}{\bibfnamefont{O.}~\bibnamefont{Morimatsu}},
  \bibnamefont{and}
  \bibinfo{author}{\bibfnamefont{T.}~\bibnamefont{Nishikawa}},
  \bibinfo{journal}{Phys. Rev.} \textbf{\bibinfo{volume}{C71}},
  \bibinfo{pages}{045202} (\bibinfo{year}{2005}), \eprint{hep-ph/0404144}.

\bibitem[{\citenamefont{Takeuchi and Shimizu}(2005)}]{Takeuchi:2004fv}
\bibinfo{author}{\bibfnamefont{S.}~\bibnamefont{Takeuchi}} \bibnamefont{and}
  \bibinfo{author}{\bibfnamefont{K.}~\bibnamefont{Shimizu}},
  \bibinfo{journal}{Phys. Rev.} \textbf{\bibinfo{volume}{C71}},
  \bibinfo{pages}{062202} (\bibinfo{year}{2005}), \eprint{hep-ph/0410286}.

\bibitem[{\citenamefont{Stancu}(2004)}]{Stancu:2004du}
\bibinfo{author}{\bibfnamefont{F.}~\bibnamefont{Stancu}},
  \bibinfo{journal}{Phys. Lett.} \textbf{\bibinfo{volume}{B595}},
  \bibinfo{pages}{269} (\bibinfo{year}{2004}), \eprint{hep-ph/0402044}.

\bibitem[{\citenamefont{Hiyama et~al.}(2005)\citenamefont{Hiyama, Kamimura,
  Hosaka, Toki, and Yahiro}}]{Hiyama:2005cf}
\bibinfo{author}{\bibfnamefont{E.}~\bibnamefont{Hiyama}},
  \bibinfo{author}{\bibfnamefont{M.}~\bibnamefont{Kamimura}},
  \bibinfo{author}{\bibfnamefont{A.}~\bibnamefont{Hosaka}},
  \bibinfo{author}{\bibfnamefont{H.}~\bibnamefont{Toki}}, \bibnamefont{and}
  \bibinfo{author}{\bibfnamefont{M.}~\bibnamefont{Yahiro}}
  (\bibinfo{year}{2005}), \eprint{hep-ph/0507105}.

\bibitem[{\citenamefont{Dmitrasinovic}(2005)}]{Dmitrasinovic:2005gq}
\bibinfo{author}{\bibfnamefont{V.}~\bibnamefont{Dmitrasinovic}},
  \bibinfo{journal}{Phys. Rev.} \textbf{\bibinfo{volume}{D71}},
  \bibinfo{pages}{094003} (\bibinfo{year}{2005}).

\bibitem[{\citenamefont{Sugiyama et~al.}(2004)\citenamefont{Sugiyama, Doi, and
  Oka}}]{Sugiyama:2003zk}
\bibinfo{author}{\bibfnamefont{J.}~\bibnamefont{Sugiyama}},
  \bibinfo{author}{\bibfnamefont{T.}~\bibnamefont{Doi}}, \bibnamefont{and}
  \bibinfo{author}{\bibfnamefont{M.}~\bibnamefont{Oka}},
  \bibinfo{journal}{Phys. Lett.} \textbf{\bibinfo{volume}{B581}},
  \bibinfo{pages}{167} (\bibinfo{year}{2004}), \eprint{hep-ph/0309271}.

\bibitem[{\citenamefont{Nishikawa et~al.}(2005)\citenamefont{Nishikawa,
  Kanada-En'yo, Morimatsu, and Kondo}}]{Nishikawa:2004tk}
\bibinfo{author}{\bibfnamefont{T.}~\bibnamefont{Nishikawa}},
  \bibinfo{author}{\bibfnamefont{Y.}~\bibnamefont{Kanada-En'yo}},
  \bibinfo{author}{\bibfnamefont{O.}~\bibnamefont{Morimatsu}},
  \bibnamefont{and} \bibinfo{author}{\bibfnamefont{Y.}~\bibnamefont{Kondo}},
  \bibinfo{journal}{Phys. Rev.} \textbf{\bibinfo{volume}{D71}},
  \bibinfo{pages}{076004} (\bibinfo{year}{2005}), \eprint{hep-ph/0411224}.

\bibitem[{\citenamefont{Kondo et~al.}(2005)\citenamefont{Kondo, Morimatsu, and
  Nishikawa}}]{Kondo:2004cr}
\bibinfo{author}{\bibfnamefont{Y.}~\bibnamefont{Kondo}},
  \bibinfo{author}{\bibfnamefont{O.}~\bibnamefont{Morimatsu}},
  \bibnamefont{and}
  \bibinfo{author}{\bibfnamefont{T.}~\bibnamefont{Nishikawa}},
  \bibinfo{journal}{Phys. Lett.} \textbf{\bibinfo{volume}{B611}},
  \bibinfo{pages}{93} (\bibinfo{year}{2005}), \eprint{hep-ph/0404285}.

\bibitem[{\citenamefont{Wang et~al.}(2005{\natexlab{a}})\citenamefont{Wang,
  Yang, and Wan}}]{Wang:2005pq}
\bibinfo{author}{\bibfnamefont{Z.-G.} \bibnamefont{Wang}},
  \bibinfo{author}{\bibfnamefont{W.-M.} \bibnamefont{Yang}}, \bibnamefont{and}
  \bibinfo{author}{\bibfnamefont{S.-L.} \bibnamefont{Wan}}
  (\bibinfo{year}{2005}{\natexlab{a}}), \eprint{hep-ph/0501015}.

\bibitem[{\citenamefont{Sasaki}(2004)}]{Sasaki:2003gi}
\bibinfo{author}{\bibfnamefont{S.}~\bibnamefont{Sasaki}},
  \bibinfo{journal}{Phys. Rev. Lett.} \textbf{\bibinfo{volume}{93}},
  \bibinfo{pages}{152001} (\bibinfo{year}{2004}), \eprint{hep-lat/0310014}.

\bibitem[{\citenamefont{Csikor et~al.}(2003)\citenamefont{Csikor, Fodor, Katz,
  and Kovacs}}]{Csikor:2003ng}
\bibinfo{author}{\bibfnamefont{F.}~\bibnamefont{Csikor}},
  \bibinfo{author}{\bibfnamefont{Z.}~\bibnamefont{Fodor}},
  \bibinfo{author}{\bibfnamefont{S.~D.} \bibnamefont{Katz}}, \bibnamefont{and}
  \bibinfo{author}{\bibfnamefont{T.~G.} \bibnamefont{Kovacs}},
  \bibinfo{journal}{JHEP} \textbf{\bibinfo{volume}{11}}, \bibinfo{pages}{070}
  (\bibinfo{year}{2003}), \eprint{hep-lat/0309090}.

\bibitem[{\citenamefont{Mathur et~al.}(2004)}]{Mathur:2004jr}
\bibinfo{author}{\bibfnamefont{N.}~\bibnamefont{Mathur}} \bibnamefont{et~al.},
  \bibinfo{journal}{Phys. Rev.} \textbf{\bibinfo{volume}{D70}},
  \bibinfo{pages}{074508} (\bibinfo{year}{2004}), \eprint{hep-ph/0406196}.

\bibitem[{\citenamefont{Ishii et~al.}(2005{\natexlab{a}})\citenamefont{Ishii,
  Doi, Iida, Oka, Okiharu, and Suganuma}}]{Ishii:2004qe}
\bibinfo{author}{\bibfnamefont{N.}~\bibnamefont{Ishii}},
  \bibinfo{author}{\bibfnamefont{T.}~\bibnamefont{Doi}},
  \bibinfo{author}{\bibfnamefont{H.}~\bibnamefont{Iida}},
  \bibinfo{author}{\bibfnamefont{M.}~\bibnamefont{Oka}},
  \bibinfo{author}{\bibfnamefont{F.}~\bibnamefont{Okiharu}}, \bibnamefont{and}
  \bibinfo{author}{\bibfnamefont{H.}~\bibnamefont{Suganuma}},
  \bibinfo{journal}{Phys. Rev.} \textbf{\bibinfo{volume}{D71}},
  \bibinfo{pages}{034001} (\bibinfo{year}{2005}{\natexlab{a}}),
  \eprint{hep-lat/0408030}.

\bibitem[{\citenamefont{Ishii et~al.}(2005{\natexlab{b}})\citenamefont{Ishii,
  Doi, Nemoto, Oka, and Suganuma}}]{Ishii:2005vc}
\bibinfo{author}{\bibfnamefont{N.}~\bibnamefont{Ishii}},
  \bibinfo{author}{\bibfnamefont{T.}~\bibnamefont{Doi}},
  \bibinfo{author}{\bibfnamefont{Y.}~\bibnamefont{Nemoto}},
  \bibinfo{author}{\bibfnamefont{M.}~\bibnamefont{Oka}}, \bibnamefont{and}
  \bibinfo{author}{\bibfnamefont{H.}~\bibnamefont{Suganuma}}
  (\bibinfo{year}{2005}{\natexlab{b}}), \eprint{hep-lat/0506022}.

\bibitem[{\citenamefont{Takahashi et~al.}(2005)\citenamefont{Takahashi, Umeda,
  Onogi, and Kunihiro}}]{Takahashi:2005uk}
\bibinfo{author}{\bibfnamefont{T.~T.} \bibnamefont{Takahashi}},
  \bibinfo{author}{\bibfnamefont{T.}~\bibnamefont{Umeda}},
  \bibinfo{author}{\bibfnamefont{T.}~\bibnamefont{Onogi}}, \bibnamefont{and}
  \bibinfo{author}{\bibfnamefont{T.}~\bibnamefont{Kunihiro}},
  \bibinfo{journal}{Phys. Rev.} \textbf{\bibinfo{volume}{D71}},
  \bibinfo{pages}{114509} (\bibinfo{year}{2005}), \eprint{hep-lat/0503019}.

\bibitem[{\citenamefont{Lasscock et~al.}(2005)}]{Lasscock:2005kx}
\bibinfo{author}{\bibfnamefont{B.~G.} \bibnamefont{Lasscock}}
  \bibnamefont{et~al.} (\bibinfo{year}{2005}), \eprint{hep-lat/0504015}.

\bibitem[{\citenamefont{Jaffe and Jain}(2005)}]{Jaffe:2004at}
\bibinfo{author}{\bibfnamefont{R.~L.} \bibnamefont{Jaffe}} \bibnamefont{and}
  \bibinfo{author}{\bibfnamefont{A.}~\bibnamefont{Jain}},
  \bibinfo{journal}{Phys. Rev.} \textbf{\bibinfo{volume}{D71}},
  \bibinfo{pages}{034012} (\bibinfo{year}{2005}), \eprint{hep-ph/0408046}.

\bibitem[{\citenamefont{Hosaka et~al.}(2005)\citenamefont{Hosaka, Oka, and
  Shinozaki}}]{Hosaka:2004bn}
\bibinfo{author}{\bibfnamefont{A.}~\bibnamefont{Hosaka}},
  \bibinfo{author}{\bibfnamefont{M.}~\bibnamefont{Oka}}, \bibnamefont{and}
  \bibinfo{author}{\bibfnamefont{T.}~\bibnamefont{Shinozaki}},
  \bibinfo{journal}{Phys. Rev.} \textbf{\bibinfo{volume}{D71}},
  \bibinfo{pages}{074021} (\bibinfo{year}{2005}), \eprint{hep-ph/0409102}.

\bibitem[{\citenamefont{Melikhov et~al.}(2004)\citenamefont{Melikhov, Simula,
  and Stech}}]{Melikhov:2004qh}
\bibinfo{author}{\bibfnamefont{D.}~\bibnamefont{Melikhov}},
  \bibinfo{author}{\bibfnamefont{S.}~\bibnamefont{Simula}}, \bibnamefont{and}
  \bibinfo{author}{\bibfnamefont{B.}~\bibnamefont{Stech}},
  \bibinfo{journal}{Phys. Lett.} \textbf{\bibinfo{volume}{B594}},
  \bibinfo{pages}{265} (\bibinfo{year}{2004}), \eprint{hep-ph/0405037}.

\bibitem[{\citenamefont{Ioffe and Oganesian}(2004)}]{Ioffe:2004qm}
\bibinfo{author}{\bibfnamefont{B.~L.} \bibnamefont{Ioffe}} \bibnamefont{and}
  \bibinfo{author}{\bibfnamefont{A.~G.} \bibnamefont{Oganesian}},
  \bibinfo{journal}{JETP Lett.} \textbf{\bibinfo{volume}{80}},
  \bibinfo{pages}{386} (\bibinfo{year}{2004}), \eprint{hep-ph/0408152}.

\bibitem[{\citenamefont{Eidemuller et~al.}(2005)\citenamefont{Eidemuller,
  Navarra, Nielsen, and Rodrigues~da Silva}}]{Eidemuller:2005jm}
\bibinfo{author}{\bibfnamefont{M.}~\bibnamefont{Eidemuller}},
  \bibinfo{author}{\bibfnamefont{F.~S.} \bibnamefont{Navarra}},
  \bibinfo{author}{\bibfnamefont{M.}~\bibnamefont{Nielsen}}, \bibnamefont{and}
  \bibinfo{author}{\bibfnamefont{R.}~\bibnamefont{Rodrigues~da Silva}},
  \bibinfo{journal}{Phys. Rev.} \textbf{\bibinfo{volume}{D72}},
  \bibinfo{pages}{034003} (\bibinfo{year}{2005}), \eprint{hep-ph/0503193}.

\bibitem[{\citenamefont{Wang et~al.}(2005{\natexlab{b}})\citenamefont{Wang,
  Yang, and Wan}}]{Wang:2005ms}
\bibinfo{author}{\bibfnamefont{Z.-G.} \bibnamefont{Wang}},
  \bibinfo{author}{\bibfnamefont{W.-M.} \bibnamefont{Yang}}, \bibnamefont{and}
  \bibinfo{author}{\bibfnamefont{S.-L.} \bibnamefont{Wan}},
  \bibinfo{journal}{Phys. Rev.} \textbf{\bibinfo{volume}{D72}},
  \bibinfo{pages}{034012} (\bibinfo{year}{2005}{\natexlab{b}}),
  \eprint{hep-ph/0504151}.

\bibitem[{\citenamefont{Melikhov and Stech}(2005)}]{Melikhov:2004ws}
\bibinfo{author}{\bibfnamefont{D.}~\bibnamefont{Melikhov}} \bibnamefont{and}
  \bibinfo{author}{\bibfnamefont{B.}~\bibnamefont{Stech}},
  \bibinfo{journal}{Phys. Lett.} \textbf{\bibinfo{volume}{B608}},
  \bibinfo{pages}{59} (\bibinfo{year}{2005}), \eprint{hep-ph/0409015}.

\bibitem[{\citenamefont{Belavin et~al.}(1975)\citenamefont{Belavin, Polyakov,
  Shvarts, and Tyupkin}}]{Belavin:1975fg}
\bibinfo{author}{\bibfnamefont{A.~A.} \bibnamefont{Belavin}},
  \bibinfo{author}{\bibfnamefont{A.~M.} \bibnamefont{Polyakov}},
  \bibinfo{author}{\bibfnamefont{A.~S.} \bibnamefont{Shvarts}},
  \bibnamefont{and} \bibinfo{author}{\bibfnamefont{Y.~S.}
  \bibnamefont{Tyupkin}}, \bibinfo{journal}{Phys. Lett.}
  \textbf{\bibinfo{volume}{B59}}, \bibinfo{pages}{85} (\bibinfo{year}{1975}).

\bibitem[{\citenamefont{Weinberg}(1975)}]{Weinberg:1975ui}
\bibinfo{author}{\bibfnamefont{S.}~\bibnamefont{Weinberg}},
  \bibinfo{journal}{Phys. Rev.} \textbf{\bibinfo{volume}{D11}},
  \bibinfo{pages}{3583} (\bibinfo{year}{1975}).

\bibitem[{\citenamefont{'t~Hooft}(1976{\natexlab{a}})}]{'tHooft:1976fv}
\bibinfo{author}{\bibfnamefont{G.}~\bibnamefont{'t~Hooft}},
  \bibinfo{journal}{Phys. Rev.} \textbf{\bibinfo{volume}{D14}},
  \bibinfo{pages}{3432} (\bibinfo{year}{1976}{\natexlab{a}}).

\bibitem[{\citenamefont{'t~Hooft}(1976{\natexlab{b}})}]{'tHooft:1976up}
\bibinfo{author}{\bibfnamefont{G.}~\bibnamefont{'t~Hooft}},
  \bibinfo{journal}{Phys. Rev. Lett.} \textbf{\bibinfo{volume}{37}},
  \bibinfo{pages}{8} (\bibinfo{year}{1976}{\natexlab{b}}).

\bibitem[{\citenamefont{Shifman et~al.}(1980)\citenamefont{Shifman, Vainshtein,
  and Zakharov}}]{Shifman:1979uw}
\bibinfo{author}{\bibfnamefont{M.~A.} \bibnamefont{Shifman}},
  \bibinfo{author}{\bibfnamefont{A.~I.} \bibnamefont{Vainshtein}},
  \bibnamefont{and} \bibinfo{author}{\bibfnamefont{V.~I.}
  \bibnamefont{Zakharov}}, \bibinfo{journal}{Nucl. Phys.}
  \textbf{\bibinfo{volume}{B163}}, \bibinfo{pages}{46} (\bibinfo{year}{1980}).

\bibitem[{\citenamefont{Shuryak and Zahed}(2004)}]{Shuryak:2003zi}
\bibinfo{author}{\bibfnamefont{E.}~\bibnamefont{Shuryak}} \bibnamefont{and}
  \bibinfo{author}{\bibfnamefont{I.}~\bibnamefont{Zahed}},
  \bibinfo{journal}{Phys. Lett.} \textbf{\bibinfo{volume}{B589}},
  \bibinfo{pages}{21} (\bibinfo{year}{2004}), \eprint{hep-ph/0310270}.

\bibitem[{\citenamefont{Isgur and Karl}(1978)}]{Isgur:1978xj}
\bibinfo{author}{\bibfnamefont{N.}~\bibnamefont{Isgur}} \bibnamefont{and}
  \bibinfo{author}{\bibfnamefont{G.}~\bibnamefont{Karl}},
  \bibinfo{journal}{Phys. Rev.} \textbf{\bibinfo{volume}{D18}},
  \bibinfo{pages}{4187} (\bibinfo{year}{1978}).

\bibitem[{\citenamefont{Okiharu et~al.}(2005)\citenamefont{Okiharu, Suganuma,
  and Takahashi}}]{Okiharu:2004wy}
\bibinfo{author}{\bibfnamefont{F.}~\bibnamefont{Okiharu}},
  \bibinfo{author}{\bibfnamefont{H.}~\bibnamefont{Suganuma}}, \bibnamefont{and}
  \bibinfo{author}{\bibfnamefont{T.~T.} \bibnamefont{Takahashi}},
  \bibinfo{journal}{Phys. Rev. Lett.} \textbf{\bibinfo{volume}{94}},
  \bibinfo{pages}{192001} (\bibinfo{year}{2005}), \eprint{hep-lat/0407001}.

\bibitem[{\citenamefont{Isgur and Karl}(1979)}]{Isgur:1978wd}
\bibinfo{author}{\bibfnamefont{N.}~\bibnamefont{Isgur}} \bibnamefont{and}
  \bibinfo{author}{\bibfnamefont{G.}~\bibnamefont{Karl}},
  \bibinfo{journal}{Phys. Rev.} \textbf{\bibinfo{volume}{D19}},
  \bibinfo{pages}{2653} (\bibinfo{year}{1979}).

\bibitem[{\citenamefont{De~Rujula et~al.}(1975)\citenamefont{De~Rujula, Georgi,
  and Glashow}}]{DeRujula:1975ge}
\bibinfo{author}{\bibfnamefont{A.}~\bibnamefont{De~Rujula}},
  \bibinfo{author}{\bibfnamefont{H.}~\bibnamefont{Georgi}}, \bibnamefont{and}
  \bibinfo{author}{\bibfnamefont{S.~L.} \bibnamefont{Glashow}},
  \bibinfo{journal}{Phys. Rev.} \textbf{\bibinfo{volume}{D12}},
  \bibinfo{pages}{147} (\bibinfo{year}{1975}).

\bibitem[{\citenamefont{Oka and Takeuchi}(1989)}]{Oka:1989ud}
\bibinfo{author}{\bibfnamefont{M.}~\bibnamefont{Oka}} \bibnamefont{and}
  \bibinfo{author}{\bibfnamefont{S.}~\bibnamefont{Takeuchi}},
  \bibinfo{journal}{Phys. Rev. Lett.} \textbf{\bibinfo{volume}{63}},
  \bibinfo{pages}{1780} (\bibinfo{year}{1989}).

\bibitem[{\citenamefont{Oka and Takeuchi}(1991)}]{Oka:1990vx}
\bibinfo{author}{\bibfnamefont{M.}~\bibnamefont{Oka}} \bibnamefont{and}
  \bibinfo{author}{\bibfnamefont{S.}~\bibnamefont{Takeuchi}},
  \bibinfo{journal}{Nucl. Phys.} \textbf{\bibinfo{volume}{A524}},
  \bibinfo{pages}{649} (\bibinfo{year}{1991}).

\bibitem[{\citenamefont{Takeuchi and Oka}(1991)}]{Takeuchi:1990qj}
\bibinfo{author}{\bibfnamefont{S.}~\bibnamefont{Takeuchi}} \bibnamefont{and}
  \bibinfo{author}{\bibfnamefont{M.}~\bibnamefont{Oka}},
  \bibinfo{journal}{Phys. Rev. Lett.} \textbf{\bibinfo{volume}{66}},
  \bibinfo{pages}{1271} (\bibinfo{year}{1991}).

\bibitem[{\citenamefont{Takeuchi and Oka}(1992)}]{Takeuchi:1992sg}
\bibinfo{author}{\bibfnamefont{S.}~\bibnamefont{Takeuchi}} \bibnamefont{and}
  \bibinfo{author}{\bibfnamefont{M.}~\bibnamefont{Oka}},
  \bibinfo{journal}{Nucl. Phys.} \textbf{\bibinfo{volume}{A547}},
  \bibinfo{pages}{283c} (\bibinfo{year}{1992}).

\bibitem[{\citenamefont{Takeuchi et~al.}(1993)\citenamefont{Takeuchi, Kubodera,
  and Nussinov}}]{Takeuchi:1993rs}
\bibinfo{author}{\bibfnamefont{S.}~\bibnamefont{Takeuchi}},
  \bibinfo{author}{\bibfnamefont{K.}~\bibnamefont{Kubodera}}, \bibnamefont{and}
  \bibinfo{author}{\bibfnamefont{S.}~\bibnamefont{Nussinov}},
  \bibinfo{journal}{Phys. Lett.} \textbf{\bibinfo{volume}{B318}},
  \bibinfo{pages}{1} (\bibinfo{year}{1993}).

\bibitem[{\citenamefont{Shinozaki et~al.}(2005)\citenamefont{Shinozaki, Oka,
  and Takeuchi}}]{Shinozaki:2004bp}
\bibinfo{author}{\bibfnamefont{T.}~\bibnamefont{Shinozaki}},
  \bibinfo{author}{\bibfnamefont{M.}~\bibnamefont{Oka}}, \bibnamefont{and}
  \bibinfo{author}{\bibfnamefont{S.}~\bibnamefont{Takeuchi}},
  \bibinfo{journal}{Phys. Rev.} \textbf{\bibinfo{volume}{D71}},
  \bibinfo{pages}{074025} (\bibinfo{year}{2005}), \eprint{hep-ph/0409103}.

\bibitem[{\citenamefont{Shuryak and Rosner}(1989)}]{Shuryak:1988bf}
\bibinfo{author}{\bibfnamefont{E.~V.} \bibnamefont{Shuryak}} \bibnamefont{and}
  \bibinfo{author}{\bibfnamefont{J.~L.} \bibnamefont{Rosner}},
  \bibinfo{journal}{Phys. Lett.} \textbf{\bibinfo{volume}{B218}},
  \bibinfo{pages}{72} (\bibinfo{year}{1989}).

\bibitem[{\citenamefont{Kochelev}(1985)}]{Kochelev:1985de}
\bibinfo{author}{\bibfnamefont{N.~I.} \bibnamefont{Kochelev}},
  \bibinfo{journal}{Yad. Fiz.} \textbf{\bibinfo{volume}{41}},
  \bibinfo{pages}{456} (\bibinfo{year}{1985}).

\bibitem[{\citenamefont{Hatsuda and Kunihiro}(1994)}]{Hatsuda:1994pi}
\bibinfo{author}{\bibfnamefont{T.}~\bibnamefont{Hatsuda}} \bibnamefont{and}
  \bibinfo{author}{\bibfnamefont{T.}~\bibnamefont{Kunihiro}},
  \bibinfo{journal}{Phys. Rept.} \textbf{\bibinfo{volume}{247}},
  \bibinfo{pages}{221} (\bibinfo{year}{1994}), \eprint{hep-ph/9401310}.

\bibitem[{\citenamefont{Naito et~al.}(2000)\citenamefont{Naito, Nemoto,
  Takizawa, Yoshida, and Oka}}]{Naito:1999sb}
\bibinfo{author}{\bibfnamefont{K.}~\bibnamefont{Naito}},
  \bibinfo{author}{\bibfnamefont{Y.}~\bibnamefont{Nemoto}},
  \bibinfo{author}{\bibfnamefont{M.}~\bibnamefont{Takizawa}},
  \bibinfo{author}{\bibfnamefont{K.}~\bibnamefont{Yoshida}}, \bibnamefont{and}
  \bibinfo{author}{\bibfnamefont{M.}~\bibnamefont{Oka}},
  \bibinfo{journal}{Phys. Rev.} \textbf{\bibinfo{volume}{C61}},
  \bibinfo{pages}{065201} (\bibinfo{year}{2000}), \eprint{hep-ph/9908332}.

\bibitem[{\citenamefont{Blask et~al.}(1990)\citenamefont{Blask, Bohn, Huber,
  Metsch, and Petry}}]{Blask:1990ez}
\bibinfo{author}{\bibfnamefont{W.~H.} \bibnamefont{Blask}},
  \bibinfo{author}{\bibfnamefont{U.}~\bibnamefont{Bohn}},
  \bibinfo{author}{\bibfnamefont{M.~G.} \bibnamefont{Huber}},
  \bibinfo{author}{\bibfnamefont{B.~C.} \bibnamefont{Metsch}},
  \bibnamefont{and} \bibinfo{author}{\bibfnamefont{H.~R.} \bibnamefont{Petry}},
  \bibinfo{journal}{Z. Phys.} \textbf{\bibinfo{volume}{A337}},
  \bibinfo{pages}{327} (\bibinfo{year}{1990}).

\bibitem[{\citenamefont{Jaffe}(1977{\natexlab{a}})}]{Jaffe:1976ig}
\bibinfo{author}{\bibfnamefont{R.~L.} \bibnamefont{Jaffe}},
  \bibinfo{journal}{Phys. Rev.} \textbf{\bibinfo{volume}{D15}},
  \bibinfo{pages}{267} (\bibinfo{year}{1977}{\natexlab{a}}).

\bibitem[{\citenamefont{Jaffe}(1977{\natexlab{b}})}]{Jaffe:1976ih}
\bibinfo{author}{\bibfnamefont{R.~L.} \bibnamefont{Jaffe}},
  \bibinfo{journal}{Phys. Rev.} \textbf{\bibinfo{volume}{D15}},
  \bibinfo{pages}{281} (\bibinfo{year}{1977}{\natexlab{b}}).

\bibitem[{\citenamefont{Jaffe}(2005)}]{Jaffe:2004ph}
\bibinfo{author}{\bibfnamefont{R.~L.} \bibnamefont{Jaffe}},
  \bibinfo{journal}{Phys. Rept.} \textbf{\bibinfo{volume}{409}},
  \bibinfo{pages}{1} (\bibinfo{year}{2005}), \eprint{hep-ph/0409065}.

\bibitem[{\citenamefont{Carlson et~al.}(2004)\citenamefont{Carlson, Carone,
  Kwee, and Nazaryan}}]{Carlson:2003xb}
\bibinfo{author}{\bibfnamefont{C.~E.} \bibnamefont{Carlson}},
  \bibinfo{author}{\bibfnamefont{C.~D.} \bibnamefont{Carone}},
  \bibinfo{author}{\bibfnamefont{H.~J.} \bibnamefont{Kwee}}, \bibnamefont{and}
  \bibinfo{author}{\bibfnamefont{V.}~\bibnamefont{Nazaryan}},
  \bibinfo{journal}{Phys. Rev.} \textbf{\bibinfo{volume}{D70}},
  \bibinfo{pages}{037501} (\bibinfo{year}{2004}), \eprint{hep-ph/0312325}.

\bibitem[{\citenamefont{Karliner and Lipkin}(2004)}]{Karliner:2004qw}
\bibinfo{author}{\bibfnamefont{M.}~\bibnamefont{Karliner}} \bibnamefont{and}
  \bibinfo{author}{\bibfnamefont{H.~J.} \bibnamefont{Lipkin}},
  \bibinfo{journal}{Phys. Lett.} \textbf{\bibinfo{volume}{B586}},
  \bibinfo{pages}{303} (\bibinfo{year}{2004}), \eprint{hep-ph/0401072}.

\end{thebibliography}

\end{document}